\documentclass[twoside]{article}
\usepackage{fleqn,espcrc2,amssymb}

\usepackage{graphicx}

\newcommand\psibar{\mathord{\overline \psi}}
\newcommand\vev[1]{\ensuremath{\left\langle#1\right\rangle}}
\def\deriv#1#{\doderiv{#1}}
\newcommand\doderiv[3]{\frac{\partial#1{#2}}{{\partial{#3}}#1}}
\newcommand\slashnext[1]{\mathpalette{\bgroup\let\style=}
                                     {\setbox0=\hbox{$\style #1$}%
                                      \setbox2=\hbox to\wd0{\hss$\style/$\hss}%
                                      \wd2=0pt\dp2=0pt\box2\box0\egroup}}
\newcommand\invisibletimes{\mathbin{}}
\newcommand\overleft[1]{\mathord{\mathop{#1}\limits^\leftarrow}}
\newcommand\overright[1]{\mathord{\mathop{#1}\limits^\rightarrow}}
\newcommand\Tr{\mathop{\rm Tr}}
\renewcommand\Im{\mathop{\rm Im}}
\renewcommand\Re{\mathop{\rm Re}}
\ifx\inlinecite\undefined\let\inlinecite=\cite\fi

\title {\hbox to \hsize{Renormalization Constants using Quark States
        in Fixed Gauge%
        \hfill\vbox to 0pt{\vss\rm\normalsize
              \hbox{LA-UR-01-0355}\vskip3\baselineskip}}}

\author{T. Bhattacharya\address{Theoretical Division, Los Alamos
                                National Laboratory, Los Alamos, New
                                Mexico 87545, USA},
        R. Gupta\addressmark,
        W. Lee\addressmark}

\begin{document}

\begin{abstract}
We present a status report on our calculation of the renormalization
constants for the quark bilinears in quenched \(O(a)\) improved Wilson
theory at \(\beta=6.4\) using quark states in Landau gauge.
\vspace{1pc}
\end{abstract}

\thispagestyle{empty}
\maketitle

\section{INTRODUCTION}
\label{sec:intro}
One of the leading uncertainties in current lattice calculations comes
from the renormalization constants necessary to relate the lattice
currents to those in some continuum renormalization scheme.  Until
recently, the most commonly used method for determining these was
1-loop perturbation theory. In the last few years a non-perturbative
method based on axial and vector Ward identities has been
developed~\cite{WI}. This allows the determination of all scale
invariant renormalization constants, and all $O(a)$ improvement
constants for bilinear operators. For the remaining bilinears and four
fermion operators that arise in the weak effective Hamiltonian, the
method of choice uses external quark states in Landau
gauge~\cite{RI}.  This defines the renormalization constants in the RI
or MOM scheme, in which the value of the renormalized operators is
specified at some fixed momentum for the external quarks. 

In this talk we present a status report of our calculations using
quark states in Landau Gauge at $\beta=6.4$.

The calculation involves two quantities: the quark propagator, \(S(p)\) and the three point
functions, \(C^\Gamma(p,q)\), 
\begin{eqnarray}
S(p)          &=& \sum_y  \vev{\psi(y) \psibar(0)} \times e^{i p y} 
\label{eq:S}
\\
C^\Gamma(p,q) &=& \sum_{xy} \vev{\psi(y) \psibar(0) \Gamma \psi(0) \psibar(x)} 
	\nonumber\\&&\qquad
                  \times e^{i p (x-y)} e^{-i q (x+y)/2} \,,
\label{eq:cgamma}
\end{eqnarray}
where \(\Gamma\) is an element of the Clifford algebra, 
\(\psi\) is the fermion field, and we only consider flavor non-singlet
operators.   In terms of these we define 
\begin{eqnarray}
\Gamma^\Sigma &=& \Im \deriv{S(p)^{-1}}{\slashnext p} 
\label{eq:truncwav}\\
\Gamma^m &=& \Re \deriv{S(p)^{-1}}{m} 
\label{eq:truncmass}\\
\Gamma^\Gamma &=& S(p)^{-1} C^\Gamma(p,0) S(p)^{-1}\,.
\label{eq:truncgam}
\end{eqnarray}
The RI scheme then corresponds to choosing a momentum $p_\mu$ 
(and mass \(m = 0\)) at which 
the renormalized inverse propagator and the renormalized truncated
three point functions have their tree level values: 
\begin{eqnarray}
\Gamma^\Sigma_R &\equiv& Z_\psi^{-1} \Gamma^\Sigma \qquad = 1
\label{eq:rencondwav}
\\
\Gamma^m_R &\equiv& Z_\psi^{-1} Z_m^{-1} \Gamma^m = 1
\label{eq:rencondmass}
\\
\Gamma^\Gamma_R &\equiv& Z_\psi^{-1} Z_\Gamma \Gamma^\Gamma
 \ \ \, = \Gamma\,.
\label{eq:rencondgam}
\end{eqnarray}
These equations, thus, define the wave function renormalization constant, \(Z_\psi\), the
mass renormalization constant, \(Z_m\), and the renormalization
constants for the bilinears, \(Z_\Gamma\). 
The resulting renormalized quantities are then also the same those in the 
continuum RI scheme at renormalization scale \(\mu^2 = p^2\). 

The most common scheme in which phenomenological results are presented
is the $\overline{MS}$, which can be obtained from the RI scheme
results by a perturbative calculation in the continuum.  However, it
is worth noting that the RI scheme does not respect the axial ward
identities, and this breaking of the ward identity cannot be accounted
for in the connection between RI and $\overline{MS}$ using
perturbation theory~~\cite{RI}.  Specifically, these identities
require that
\begin{eqnarray}
\gamma_5\invisibletimes (1 + m_R \deriv{}{m_R}) \Gamma^P_R =
   -\Gamma^m_R \\ 
\gamma_5\invisibletimes (1 - q_\mu \deriv{}{q_\mu}) \Gamma^A_R =
   -\Gamma^\Sigma_R 
\end{eqnarray}
at \(q_\mu = m = 0\), where \(m_R = Z_m m\) is the renormalized mass
and $P$ and $A$ are the pseudoscalar and axial bilinears.
Fortunately, the derivative terms in these equations, which contribute
due to the Goldstone pole at zero quark mass arising out of chiral
symmetry breaking, become irrelevant when the renormalization scale
\(\mu\) is large compared to the scale of chiral symmetry
breaking~\cite{RI}.  On the other hand lattice discretization errors
become large for $pa$ larger than unity. Thus, the method relies on
the existence of an intermediate window in which these two artifacts
are small.

\section{DISCRETIZATION EFFECTS}
\label{sec:discr}
In the lattice regularization scheme operators receive power
corrections which involve higher dimension operators. This is a
consequence of the hard cutoff, $i.e.$ the lattice spacing \(a\). In
addition, because the RI scheme is defined in a fixed gauge,
gauge-invariant and gauge-dependent operators, in general, mix.
Application of BRS symmetry shows~\cite{ROME:Imp:97} that, at \(O(a)\) and \(q=0\), the
only corrections to \(\psibar \Gamma \psi\) are gauge invariant and of
the form \(\psibar (\overleft {\cal W} \Gamma + \Gamma \overright
{\cal W}) \psi\), where \({\cal W}\) is the Dirac operator and this
correction vanishes by the equation of motion, \( \psibar \overleft
{\cal W} = \overright {\cal W} \psi = 0 \).  Such a correction term,
therefore, does not affect the on-shell matrix elements of these
operators, but does contribute to the momentum space correlators in
Eq.~\ref{eq:cgamma} as these involve contributions from points where
the operator and the fermion sources overlap.

At $O(a)$ only $ \slashnext {\overright D} - m $ part of $ \overright
{\cal W} $ contributes to the mixing; consequently, the scalar
density and the vector current mix, as do the axial current and the
tensor density. The pseudoscalar current receives only multiplicative
corrections at this order.  At higher orders, the violation of
rotational and Lorentz symmetries show up, and different components of
\(V_\mu\), \(A_\mu\), and \(T_{\mu\nu}\) mix amongst themselves.

To take the $O(a)$ mixing into account, Eq.~\ref{eq:rencondgam} is 
modified to 
\begin{eqnarray}
\Gamma^\Gamma_R &\equiv& Z_\psi^{-1} Z_\Gamma \Gamma^\Gamma_I \\
\Gamma^\Gamma_I &\equiv& \Gamma^\Gamma +
    c'_\Gamma (\Gamma S^{-1} + S^{-1} \Gamma)\,,
\label{eq:improve}
\end{eqnarray}
where \(c'_\Gamma\) is determined by the requirement that 
\begin{eqnarray}
\Tr \Gamma^\Gamma_I \Gamma' = 0 \,,
\label{eq:cprime}
\end{eqnarray}
where $\Gamma' = S$, $V$, $T$, and $A$ for $\Gamma = V$, $S$, $A$, and
$T$ respectively.  Unfortunately, as stated above, there is no
analogous condition to determine \(c'_P\) since $\gamma_5 S^{-1} +
S^{-1} \gamma_5 \propto \gamma_5$. An illustration of the magnitude of
the mixing is given in Table~\ref{tab:table1} from which the
$c_\Gamma'$ are determined.  Similarly, at \(O(a)\), the propagator
gets corrections
\begin{eqnarray}
S_I (p) = (1 - 2 a c^{NGI}_\psi \slashnext p) S (p) - 2 a c'_\psi \,,
\end{eqnarray}
where \(c^{NGI}_\psi\) and \(c'_\psi\) are the gauge-variant and
gauge-invariant corrections respectively.  Methods to determine a
combination of these coefficients have been discussed in
Ref.~\inlinecite{qimp} and we do not repeat the discussion here.

\section{LATTICE PARAMETERS AND METHODOLOGY}
Our pilot calculation utilizes 60 lattices at \(\beta = 6.4\) which
corresponds to a lattice scale of \(a^{-1} \approx 3.85 {\rm
GeV}\)~\cite{Schierholz} and a critical hopping parameter \(\kappa_c
\approx 0.135796\)~\cite{Schiercalc}.  The quark
propagators were inverted at seven values of \(\kappa = 0.1280\),
0.1294, 0.1308, 0.1324, 0.1334, 0.1343, and 0.1348 using a clover
coefficient value of \(c_{SW} = 1.526\)~\cite{Luescher}.  Unless stated
otherwise, we shall only consider momenta such that all components
are less than \(\pi/4\).\looseness-1

The truncated three-point functions were calculated and projected on
the Clifford basis.  As shown in Tab.~\ref{tab:table1}, the data, in
addition to the expected terms, shows $O(a)$ mixing ($S
\leftrightarrow V_i$), and the two $O(a^2)$ effects: mixing ($V_3
\leftrightarrow V_4$) and the difference between the terms parallel
and perpendicular to the momentum.  The latter effect is due to the
violation of rotational symmetry.  As mentioned in the previous
section, we correct for the \(O(a)\) mixings by determining the
$c'_\Gamma$, but, at \(pa \sim 1\), the approximately \(2\%\)
violation of rotational symmetry and \(0.6\%\) mixings due to higher
order effects survive.

\begin{table}[htb]
\caption{The various Clifford projections of the truncated 3-point
functions corresponding to the scalar and vector bilinears at
\(\kappa=0.1294\) and \(pa = (0,0,3,3)\times \pi/16\).}
\begin{center}
\renewcommand{\tabcolsep}{1pc} 
\renewcommand{\arraystretch}{1.2} 
\begin{tabular}{ccl}
\hline
operator&projection&\multicolumn{1}{c}{value}\\
\hline
\( S \)&\(  1     \)&\(1.180(5)\)\\
       &\(\gamma_3\)&\(0.059(3)\)\\
       &\(\gamma_4\)&\(0.059(3)\)\\
\(V_1\)&\(\gamma_1\)&\(1.067(5)\)\\
\(V_2\)&\(\gamma_2\)&\(1.067(5)\)\\
\(V_3\)&\(\gamma_3\)&\(1.086(3)\)\\
       &\(  1     \)&\(0.019(2)\)\\
       &\(\gamma_4\)&\(0.006(1)\)\\
\(V_4\)&\(\gamma_4\)&\(1.083(3)\)\\
       &\(  1     \)&\(0.021(2)\)\\
       &\(\gamma_3\)&\(0.006(1)\)\\
\hline
\end{tabular}
\end{center}
\label{tab:table1}
\end{table}

To calculate the wave function renormalization, we Fourier transform
the inverse propagator,
\begin{equation}
S^{-1} (p) = \sum_{\{n\}} a_{\{n\}} \prod_\mu \sin n_\mu p_\mu\,,
\end{equation}
and calculate the derivative in Eq.~\ref{eq:truncwav} analytically:
\begin{equation}
\Gamma^\Sigma = \frac{1}{4} \sum_\mu \frac{1}{4} 
    \Tr \gamma_\mu \sum_{\{n\}} n_\mu a_{\{n\}} \prod_\nu \cos n_\nu p_\nu\,.
\end{equation}
We expect large \(O(a^2)\) discretization errors for the following
reason.  At tree level \( \Gamma^\Sigma = (1/4) \sum_\mu cos p_\mu\),
which deviates from unity by $\sim 15\%$ at \(pa \sim 1\). To correct
for this artifact, we divide the calculated \(\Gamma^\Sigma\) by this
tree-level contribution, and use the resulting \( \Gamma^\Sigma \) in
Eq.~\ref{eq:rencondmass} to calculate \(Z_\psi\).  Note that such a
determination of \(Z_\psi\) contains residual \(O(a)\) errors stemming
from the use of the uncorrected propagator.

\section{RESULTS}
\begin{figure}[htb]
\begin{center}
\includegraphics[width=0.9\hsize]{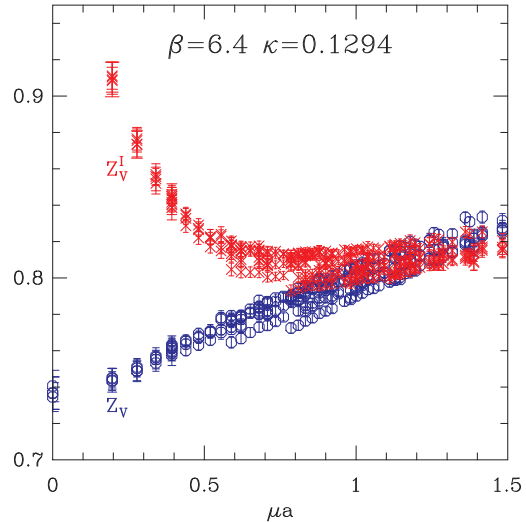}
\end{center}
\caption{The improved (crosses) and unimproved (circles) vector
renormalization constants as a function of the renormalization scale.}
\label{fig:figure1}
\end{figure}
In Fig.~\ref{fig:figure1}, we show the effect of improvement for the
vector current, $i.e.$ removing the \(O(a)\) term by calculating
$c'_V$.  In the absence of discretization errors, the calculated
\(Z_V\) should be independent of the scale \(\mu\) except for
non-perturbative terms.  Non-perturbative terms are expected to fall
off as \(1/\mu^2\) at large \(\mu\)~\cite{RI}.  Instead the
uncorrected data shows a linear rise, \(Z_V\) changing by $\sim 5\%$
between \(0.8 < \mu a < 1.5\).  In contrast, the \(O(a)\) corrected
quantity\footnote{
arising from the use of the uncorrected propagator to define
\(Z_\psi\).}  is constant over this range of \(\mu a\). There is
roughly \(2-3\%\) uncertainty in \(Z_V\) coming from the variation
between different Lorentz components and different combinations of
momenta in this range.

\begin{figure}[htb]
\begin{center}
\includegraphics[width=0.9\hsize]{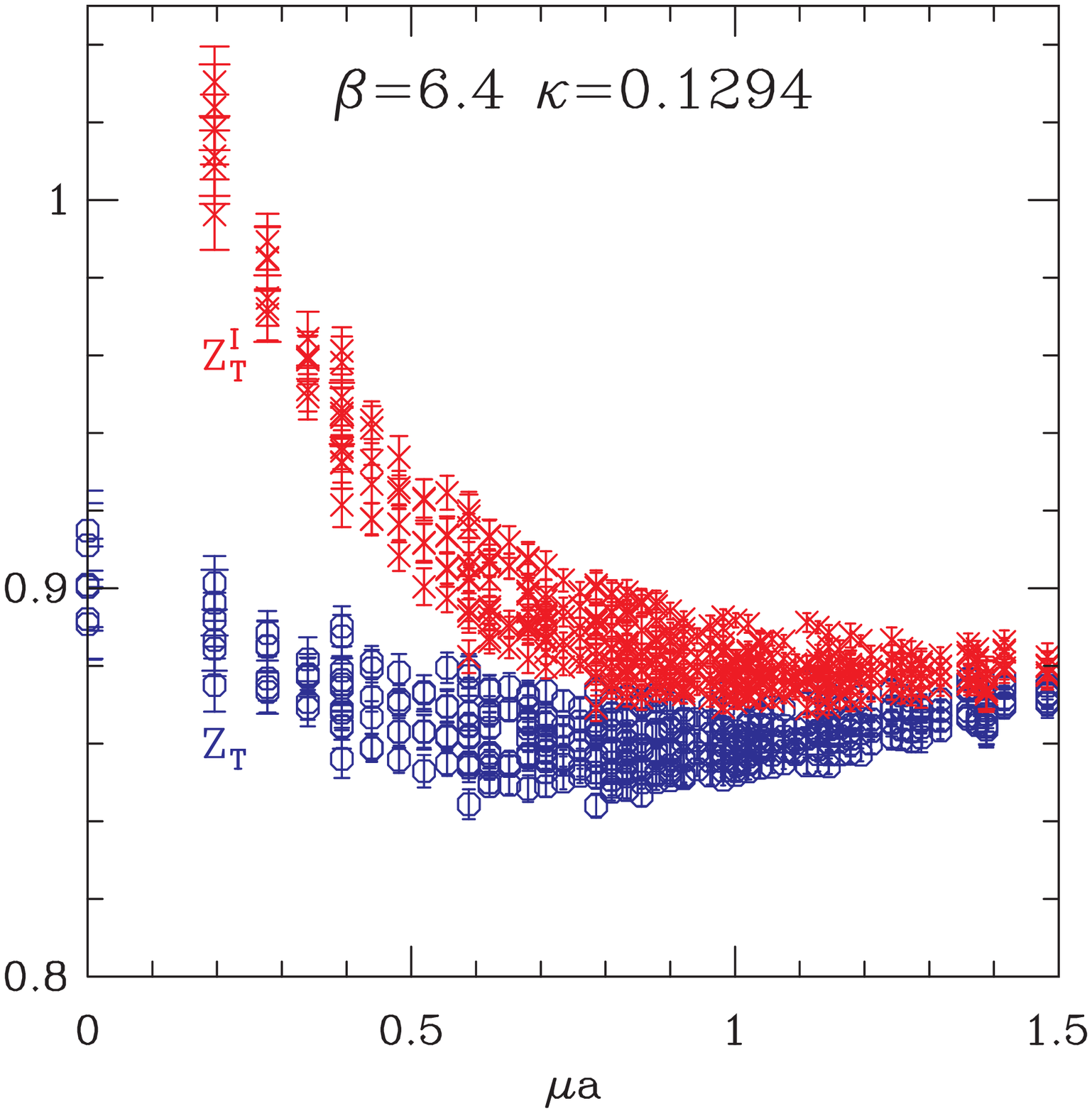}
\end{center}
\caption{The improved (crosses) and unimproved (circles) tensor
renormalization constants as a function of the renormalization scale.}
\label{fig:figure2}
\begin{center}
\includegraphics[width=0.9\hsize]{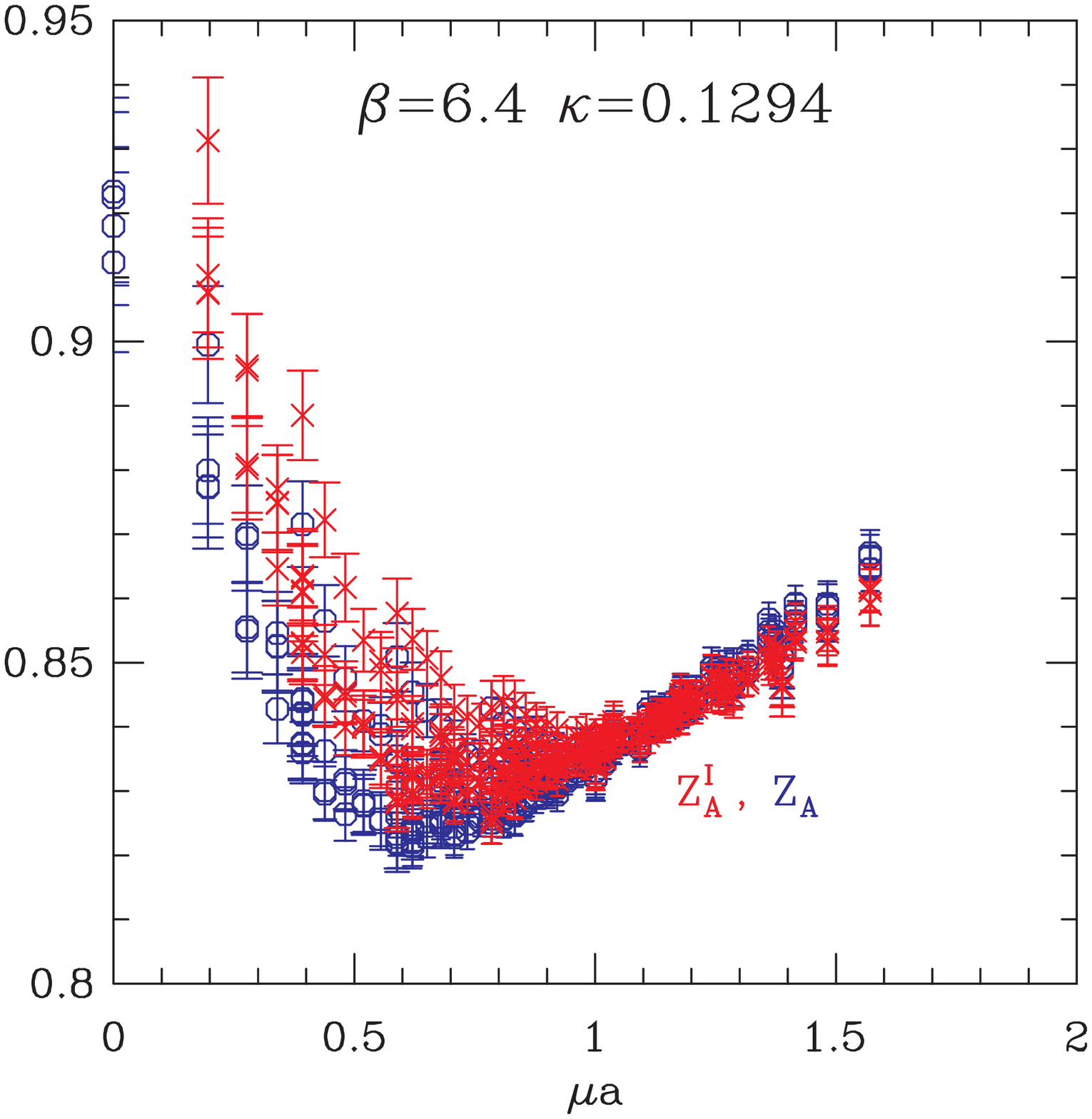}
\end{center}
\caption{The improved (crosses) and unimproved (circles) axial vector
renormalization constants as a function of the renormalization scale.}
\label{fig:figure3}
\end{figure}

Fig.~\ref{fig:figure2} shows that the behavior of tensor
renormalization constant, \(Z_T\), is qualitatively similar: the much
smaller linear rise of the unimproved $Z_T$ for \(\mu a > 0.8\) is
removed by the improvement.  Finally, we show the behavior of the
axial vector channel in Fig.~\ref{fig:figure3}. We do not observe any
improvement in $Z_A$ and the data do not show a window in which it is
independent of $\mu a$.

\section{DISCUSSIONS}
Since we have not finished the analysis, we end with a few qualitative 
statements. (i) The \(O(a)\) errors induced due to mixing with the 
equation of motion operators can be corrected for by calculating 
$c'_\Gamma$. These $c'_\Gamma$ can then be compared with those evaluated 
using the axial Ward identity~\cite{WI}. 
(ii) At \(\beta=6.4\), the presence of
non-perturbative corrections forces us to work at \(\mu a \sim 0.8\)
or higher to determine the renormalization constants. At these large
momenta, the \(O(a^2)\) errors are only slightly smaller than the
\(O(a)\) errors, and it is important to ascertain whether our attempts to
correct the latter introduce unacceptably large \(O(a^2)\) errors.
In particular, the correction terms in Eq.~\ref{eq:improve} involve
the inverse propagator, and an \(O(a)\) improvement in the propagator
used in this equation changes results at \(O(a^2)\).  The use of a
suitably improved propagator would make the $O(a)$ correction term vanish in the
chiral limit except for non-perturbative effects that fall off as
\(1/p^2\).  In practice, as already noted in Ref.~\cite{qimp},
the bare inverse propagator has large \(O(ap^2)\) corrections even in the
chiral limit, and hence our improvement term does not vanish there.

To summarize, \(O(a^2)\) corrections are large and, at present,
uncontrolled.  We are investigating ways to improve the calculation
and the results of this study will be reported elsewhere.

\end{document}